\documentclass[12pt]{article}
\usepackage[a4paper, margin=1in]{geometry}
\usepackage{graphicx}
\usepackage{amsmath}
\usepackage{amssymb}
\usepackage{natbib}
\usepackage{booktabs}
\usepackage{threeparttable}
\usepackage{pdflscape}
\usepackage{longtable}
\usepackage{setspace}
\usepackage{comment,orcidlink}
\usepackage{rotating}
\usepackage{xcolor}
\usepackage{chngcntr}

\makeatletter
\renewenvironment{abstract}
  {
   \begin{center}
   \bfseries \abstractname
   \end{center}
   \singlespacing
   \noindent
  }
  {\par}
\makeatother
\newcommand{\blind}{0}
\newcommand{\sym}[1]{\ifmmode^{#1}\else\(^{#1}\)\fi}

\begin{document}
\def\spacingset#1{\renewcommand{\baselinestretch}{#1}\small\normalsize} \spacingset{1}
\if0\blind
{
\title{\bf Market Efficiency and the Heterogeneous Impact of Financial Liberalization: Evidence from the Shanghai-Hong Kong Stock Connect}} 
\author{\normalsize Jiaqi Liu \orcidlink{0009-0001-8400-3289} \footnote{Postal address: Research School of Finance, Actuarial Studies and Statistics, Level 4, Building 26C, Australian National University, Kingsley Street, Acton, Canberra, ACT 2601, Australia; Email: U7769449@anu.edu.au.}\\
\normalsize Research School of Finance, Actuarial Studies and Statistics \\
\normalsize Australian National University \\
\\	
\normalsize Chen Tang \orcidlink{0000-0002-0948-6073}
 \\
\normalsize Research School of Finance, Actuarial Studies and Statistics \\
\normalsize Australian National University \\
\\
}
\date{}
\maketitle
\date{\vspace{-5ex}}
\maketitle
\maketitle

\begin{abstract}
\singlespacing
This paper examines how the Shanghai–Hong Kong Stock Connect (SHHK Stock Connect) affects the A–H share price premium and whether the policy effect depends on pre-existing market efficiency. Using monthly data for 67 A–H dual-listed firms from 2011 to 2019, we estimate dynamic models using system GMM. We find that the implementation of SHHK Stock Connect is associated with an average 18.4\% increase in the A–H price premium, indicating a widening rather than a narrowing of the A–H share price premium. More importantly, this effect is heterogeneous: the policy impact is more pronounced for firms operating in less efficient markets and weaker for those with higher efficiency. We find no significant pricing response at the announcement stage. These findings imply that financial liberalization does not necessarily promote cross-market price convergence, its effects depend on pre-existing market efficiency, suggesting that cross-border liberalization policies may be more effective when accompanied by improvements of market efficiency.


\end{abstract}
\newpage
\singlespacing
\section{Introduction}\label{sec:1}

Dual-listed firms are entities whose equity is simultaneously traded on a domestic exchange and one or more international exchange \citep{alexanderAssetPricingDual1987}. Empirical evidence suggests that dual-listed firms frequently exhibit significant price disparities between their domestic and international share classes \citep{baileyRiskReturnChinas1994}. An example is the persistent A-H premium \footnote{A-H premium means, for a A-H dual listing firm, the ratio of A-share price to H-share price after adjustment of foreign exchange rate.}, where domestically listed A-shares consistently trade at a higher price than their corresponding H-shares. Theoretically, based on the Law of One Price, the prices of identical assets should be the same after considering exchange rates, otherwise arbitrage opportunity exists. In this paper, we examine whether capital market liberalization affects the A-H premium and whether market efficiency conditions this effect.

On 10 April 2014, the China Securities Regulatory Commission and the Securities and Futures Commission jointly approved the framework for the SHHK Stock Connect, establishing an institutional mechanism for cross-border equity trading between the Shanghai and Hong Kong markets. The program formally commenced on 17 November 2014. Prior studies provide mixed evidence on its pricing effect: while \citet{chanCapitalAccountLiberalization2016} document a widening of the A-H premium after implementation, \citet{fanImpactShanghaiHong2017} find that the SHHK Stock Connect narrowed A-H share price differences.

Prior research has focused on the average effect of the SHHK Stock Connect, yet the findings remain inconclusive. \citet{bekaertDoesFinancialLiberalization2005} find that the effects of financial liberalization depend on pre-existing levels of financial development and institutional quality. This heterogeneity is further supported by \citet{mittonStockMarketLiberalization2006}, who demonstrates that the positive impacts of stock market opening are significantly stronger for certain firms, depending on their characteristics and degree of "investability." \citet{famaEfficientCapitalMarkets1970} proposes that share prices should immediately incorporate all available information from the market when the market is efficient. Existing studies provide limited evidence on whether the effect of the SHHK Stock Connect depends on pre-existing market efficiency. This research gap matters because financial liberalization may not generate uniform price convergence when market inefficiencies persist. Therefore, this study examines both the average effect of the SHHK Stock Connect on the A--H premium and its dependence on the efficiency of the connected markets.

Our result shows that the implementation of SHHK Stock Connect is associated with an average 18.4\% increase in the A–H premium and this effect is stronger for firms operating in less efficient markets and weaker for firms in more efficient markets. The findings are stable after conducting several robustness checks.

This paper makes two contributions: Firstly, the heterogeneous effect of SHHK Stock Connect helps explain the mixed findings in previous. Secondly, this paper highlights pre-existing market efficiency as an important condition of the pricing consequences of financial liberalization. These findings also provide implications for policymakers, regulators and investors. In particular, they suggest that financial liberalization may be more effective when accompanied by reforms that improve the information environment. For investors, the findings suggest that, although SHHK Stock Connect expands the investor pool and facilitates cross-border trading, arbitrage across markets may remain limited when the information environment is weak.

\section{Data and Methodology}\label{sec:2}
\subsection{Sample data}\label{prem}
Our sample comprises 67 Shanghai-listed A-H dual listings that completed dual-listing before the SHHK Stock Connect, observed monthly from January 2011 to May 2019. Table \ref{tab:1} presents the characteristic of these firms. The endpoint helps isolate the implementation effect before later shocks, such as the Anti-ELAB movement and COVID-19 \citep{allenDissectingLongTermPerformance2024}. Then, the objective is to identify the pricing effect of the SHHK Stock Connect, rather than to characterize the long-run evolution of the A–H premium. Our stock price data are from Wind Terminal, firm characteristic data are obtained from Choice database. Our dependent variable, the A-H premium ($\texttt{Prem}$), is defined as the exchange-rate-adjusted ratio of A-share to H-share prices \citep{fanImpactShanghaiHong2017}.

\begin{equation*}
\text{Prem}_{i,t}
= \frac{P^{A}_{i,t}}{P^{H}_{i,t}\times \text{HKD/CNY}_{t}} - 1,
\end{equation*}
The primary independent variable, $\texttt{SHHKPolicy}$, is an indicator equal to one after November 2014, otherwise zero. Following \citet{corwinSimpleWayEstimate2012} and \citet{mengEffectOverseasInvestors2023}, we proxy market efficiency using the negative value of high-low bid-ask spreads\footnote{See details in Appendix~\ref{app:B}}. Control variables include relative liquidity ($\texttt{Turnover}$), log-transformed market capitalization ($\texttt{Size}$), and a state-ownership indicator ($\texttt{SOE}$), which equals to one if the firm is state owned, otherwise zero. Furthermore, we define relative share demand ($\texttt{Demand}$) as the ratio of floating A-shares to H-shares \citep{chanMarketSegmentationShare2005} and relative monetary conditions ($\texttt{Interest\ rate}$) as the 3-months SHIBOR-to-HIBOR ratio.

\subsection{Methodology}
We employ system GMM estimator rather than traditional fixed-effect model for three main reasons: Firstly, based on \citet{domowitzMarketSegmentationStock1997}, the share price premium of dual-listed firms is persistent, requiring a dynamic model with a lagged dependent variable. Secondly, some control variables may be endogenous. Finally, given the persistence of premium, difference GMM may suffer from weak instruments, whereas system GMM improves efficiency by combining the equations in differences and levels under additional moment conditions \citep{blundellInitialConditionsMoment}. Although system GMM can improve efficiency relative to difference GMM, it may still face weak-instrument problems and inference can be sensitive \citep{bunWeakInstrumentProblem2010}.
The regression model is listed below:

\begin{equation*}
\text{Prem}_{i,t}
=
\alpha_i
+
\beta \, \text{Prem}_{i,t-1}
+
\theta \, \text{SHHKPolicy}_t
+
\gamma' X_{i,t}
+
\varepsilon_{i,t}.
\end{equation*}
where the $\text{Prem}_{i,t}$ is the A-H share price premium for share $i$ at month $t$ and $X_{i,t}$ is a vector of control variables.

\section{Empirical results}\label{sec:3}
Table \ref{tab:2} lists the descriptive statistics for main variables used in the regression model. We can see that the mean value of the premium is 68.3\%.
\subsection{Baseline Effect of SHHK Stock Connect}
We examine the effect of the SHHK Stock Connect on the A-H share price premium. Table~\ref{tab:3} reports the regression results. In column~(1), the coefficient on \texttt{SHHKPolicy} is positive and statistically significant, indicating that the policy enlarges A-H share price premium. Columns~(2) to (6) gradually incorporate additional control variables, the estimated coefficient on \texttt{SHHKPolicy} remains positive and significant at 1\% level. In column~(6), the coefficient on \texttt{SHHKPolicy} is 0.184, implying that, holding other variables constant, the implementation of the SHHK Stock Connect is associated with an average 18.4\% increase in the A-H share price premium for a given pair of dual-listed shares.
\subsection{Market Efficiency-Dependent Effects of the SHHK Stock Connect}
Cross-market frictions can impede arbitrage between A- and H-shares, potentially leading to heterogeneous pricing responses to policy shocks \citep{gagnonMultimarketTradingArbitrage2010}. When market efficiency is compromised, asset prices weaken their relationship with information-based equilibrium and become more susceptible to speculative trading. This dynamic can be described by other models that suggest speculative sentiment can spread among investor groups, much like the spread of a infectious disease \citep{aliuInfectiousDiseasesExplain2024, aliuBitcoinInfectionDisease2025}. Recent evidence suggests that the SHHK Stock Connect improves market efficiency for A-share firms \citep{mengEffectOverseasInvestors2023}. These findings raise the question of whether the pricing impact of the SHHK Stock Connect depends on pre-existing market efficiency. To examine this hypothesis, we estimate the following model:
\begin{equation*}
\text{Prem}_{i,t} = \alpha_i +\beta_{1} \text{Prem}_{i,t-1}+\beta_{2} \text{SHHKPolicy}_t+\beta_{3}\text{Effboth}_t+ \beta_{4} \text{SHHKPolicy}\times\text{Effboth}_t + \gamma' X_{i,t} + \epsilon_{i,t},
\end{equation*}
where the $\text{Prem}_{i,t}$ is the A-H share price premium for share $i$ at month $t$. 
Table~\ref{tab:4} reports the regression results. After considering the interaction term between the Stock Connect indicator and market efficiency (\texttt{SHHKPolicy}$\times$\texttt{Effboth}) and efficiency for both market (\texttt{Effboth}), the coefficient of \texttt{SHHKPolicy} becomes negative and statistically significant.
Given that $\texttt{Effboth}$ is constructed as the negative of the bid-ask spread, higher values (closer to zero) indicate higher efficiency. The marginal policy effect is $\beta_{2} + \beta_{4}\texttt{Effboth}$, implying that the impact of SHHK increases as market efficiency declines. Overall, the results in Table~\ref{tab:4} support an efficiency-dependent policy effect both statistically and economically.
\subsection{Robustness check}
We conduct several robustness checks to make result more reliable. Firstly, we examine whether the policy announcement (in April 2014\footnote{On 10 April 2014, CRSC and SFC approved SHHK Stock Connect.}) affects the A-H premium. Appendix~\ref{app:C} presents the regression results. Appendix~\ref{app:D} shows whether the announcement of the policy has a similar effect on the A-H price premium. Appendix~\ref{app:E} lists the alternative market efficiency proxy construction and regression results. Appendix~\ref{app:F} presents placebo tests by assuming the SHHK Stock Connect was implemented two years later. In Appendix~\ref{app:G}, we control extra variables. These results are consistent with the main results, indicating our findings are reliable.

\section{Conclusion and Limitation}\label{sec:4}
This paper examines whether the SHHK Stock Connect affects the A--H premium and whether this effect depends on market efficiency. We find that implementation increases the premium on average, with stronger effects among firms in less efficient markets and no significant announcement response.

These findings suggest that cross-border market liberalization alone may not produce price convergence, its effectiveness depends on pre-existing market efficiency. Therefore, liberalization policies may be more effective when accompanied by measures that improve market efficiency, for example, better ESG-related disclosure.

This study has several limitations. The sample focuses on Shanghai-listed A–H firms and ends in May 2019, which may limit the generalizability of the results to other markets and later periods. Future research may extend the sample to include additional firms and examine whether these effects remain stable under more recent market conditions.

\newpage
\subsubsection*{Author Contributions}
CRediT: \textbf{Jiaqi Liu:} Writing - original draft, Conceptualisation, Methodology, Data curation, Validation \textbf{Chen Tang:} Supervision, Writing-review \& editing
\subsubsection*{Disclosure Statement:}
No potential conflict of interest was reported by the author(s). 
\subsubsection*{Funding:}
No funding was received for this research.
\subsubsection*{Word Count: 1998} 

\subsubsection*{Generative AI tools use:} The authors used ChatGPT 5.2 and 5.3 to improve the clarity of the manuscript and to assist with coding tasks, the results are verified by the authors.

\newpage
\bibliographystyle{apalike}
\bibliography{references}

@article{alexanderAssetPricingDual1987,
  title = {Asset {{Pricing}} and {{Dual Listing}} on {{Foreign Capital Markets}}: {{A Note}}},
  shorttitle = {Asset {{Pricing}} and {{Dual Listing}} on {{Foreign Capital Markets}}},
  author = {Alexander, Gordon J. and Eun, Cheol S. and Janakiramanan, S.},
  year = 1987,
  journal = {The Journal of Finance},
  volume = {42},
  number = {1},
  eprint = {2328425},
  eprinttype = {jstor},
  pages = {151--158},
  publisher = {[American Finance Association, Wiley]},
  issn = {0022-1082},
  doi = {10.2307/2328425},
  urldate = {2026-04-10},
  jstor = {2328425}
}

@article{aliuBitcoinInfectionDisease2025,
  title = {Bitcoin as an Infection Disease: Evidence from {{SIR}} Epidemiological Model},
  shorttitle = {Bitcoin as an Infection Disease},
  author = {Aliu, Florin},
  year = 2025,
  month = nov,
  journal = {Review of Behavioral Finance},
  volume = {17},
  number = {6},
  pages = {961--978},
  issn = {1940-5979, 1940-5987},
  doi = {10.1108/RBF-05-2025-0185},
  urldate = {2026-04-16},
  language = {en}
}

@article{aliuInfectiousDiseasesExplain2024,
  title = {Do Infectious Diseases Explain {{Bitcoin}} Price {{Fluctuations}}?},
  author = {Aliu, Florin},
  year = 2024,
  month = jun,
  journal = {Journal of International Financial Markets, Institutions and Money},
  volume = {93},
  pages = {102011},
  issn = {10424431},
  doi = {10.1016/j.intfin.2024.102011},
  urldate = {2026-04-14},
  language = {en}
}

@article{allenDissectingLongTermPerformance2024,
  title = {Dissecting the {{Long-Term Performance}} of the {{Chinese Stock Market}}},
  author = {Allen, Franklin and Qian, Jun (qj) and Shan, Chenyu and Zhu, Julie Lei},
  year = 2024,
  journal = {The Journal of Finance},
  volume = {79},
  number = {2},
  pages = {993--1054},
  issn = {1540-6261},
  doi = {10.1111/jofi.13312},
  urldate = {2026-01-09},
  language = {english}
}

@article{baileyRiskReturnChinas1994,
  title = {Risk and Return on {{China}}'s New Stock Markets: {{Some}} Preliminary Evidence},
  author = {Bailey, Warren},
  year = 1994,
  journal = {Pacific-Basin Finance Journal},
  volume = {2},
  number = {2},
  pages = {243--260},
  issn = {0927-538X},
  doi = {10.1016/0927-538X(94)90019-1}
}

@article{bekaertDoesFinancialLiberalization2005,
  title = {Does Financial Liberalization Spur Growth?},
  author = {Bekaert, Geert and Harvey, Campbell R. and Lundblad, Christian},
  year = 2005,
  month = jul,
  journal = {Journal of Financial Economics},
  volume = {77},
  number = {1},
  pages = {3--55},
  issn = {0304-405X},
  doi = {10.1016/j.jfineco.2004.05.007},
  urldate = {2026-02-04}
}

@article{blundellInitialConditionsMoment,
  title = {Initial Conditions and Moment Restrictions in Dynamic Panel Data Models},
  author = {Blundell, Richard and Bond, Stephen},
  year = 1998,
  journal = {Journal of Econometrics},
  volume = {87},
  pages = {115--143},
  language = {en}
}

@article{bunWeakInstrumentProblem2010,
  title = {The Weak Instrument Problem of the System {{GMM}} Estimator in Dynamic Panel Data Models},
  author = {Bun, Maurice J. G. and Windmeijer, Frank},
  year = 2010,
  month = feb,
  journal = {The Econometrics Journal},
  volume = {13},
  number = {1},
  pages = {95--126},
  issn = {1368-4221, 1368-423X},
  doi = {10.1111/j.1368-423X.2009.00299.x},
  urldate = {2026-04-20},
  copyright = {http://doi.wiley.com/10.1002/tdm\_license\_1.1},
  language = {en}
}

@article{chanCapitalAccountLiberalization2016,
  title = {Capital Account Liberalization and Dynamic Price Discovery: {{Evidence}} from {{Chinese}} Cross-Listed Stocks},
  shorttitle = {Capital Account Liberalization and Dynamic Price Discovery},
  author = {Chan, Marc K. and Kwok, Simon S.},
  year = 2016,
  month = feb,
  journal = {Applied Economics},
  volume = {48},
  number = {6},
  pages = {517--535},
  issn = {0003-6846, 1466-4283},
  doi = {10.1080/00036846.2015.1083087},
  urldate = {2026-02-03},
  language = {english}
}

@article{changCrosslistingPricingEfficiency,
  title = {Cross-Listing and Pricing Efficiency: {{The}} Informational and Anchoring Role Played by the Reference Price},
  author = {Chang, Eric C and Luo, Yan and Ren, Jinjuan},
  year = 2013,
  journal = {Journal of Banking \& Finance},
  volume = {37},
  pages = {4449--4464},
  doi = {10.1016/j.jbankfin.2012.12.018},
  language = {en}
}

@article{chanMarketSegmentationShare2005,
  title = {Market {{Segmentation}} and {{Share Price Premium}}: {{Evidence}} from {{Chinese Stock Markets}}},
  shorttitle = {Market {{Segmentation}} and {{Share Price Premium}}},
  author = {Chan, Kalok and Kwok, Johnny K.H.},
  year = 2005,
  month = apr,
  journal = {Journal of Emerging Market Finance},
  volume = {4},
  number = {1},
  pages = {43--61},
  issn = {0972-6527, 0973-0710},
  doi = {10.1177/097265270400400103},
  urldate = {2026-01-09},
  copyright = {https://journals.sagepub.com/page/policies/text-and-data-mining-license},
  language = {en}
}

@article{corwinSimpleWayEstimate2012,
  title = {A {{Simple Way}} to {{Estimate Bid-Ask Spreads}} from {{Daily High}} and {{Low Prices}}},
  author = {Corwin, Shane A. and Schultz, Paul},
  year = 2012,
  month = apr,
  journal = {The Journal of Finance},
  volume = {67},
  number = {2},
  pages = {719--760},
  issn = {0022-1082, 1540-6261},
  doi = {10.1111/j.1540-6261.2012.01729.x},
  urldate = {2026-01-09},
  language = {en}
}

@article{domowitzMarketSegmentationStock1997,
  title = {Market {{Segmentation}} and {{Stock Prices}}: {{Evidence}} from an {{Emerging Market}}},
  shorttitle = {Market {{Segmentation}} and {{Stock Prices}}},
  author = {Domowitz, Ian and Glen, Jack and Madhavan, Ananth},
  year = 1997,
  journal = {The Journal of Finance},
  volume = {52},
  number = {3},
  pages = {1059--1085},
  issn = {1540-6261},
  doi = {10.1111/j.1540-6261.1997.tb02725.x},
  urldate = {2026-01-20},
  language = {en}
}

@article{famaEfficientCapitalMarkets1970,
  title = {Efficient {{Capital Markets}}: {{A Review}} of {{Theory}} and {{Empirical Work}}},
  shorttitle = {Efficient {{Capital Markets}}},
  author = {Fama, Eugene F.},
  year = 1970,
  journal = {The Journal of Finance},
  volume = {25},
  number = {2},
  eprint = {2325486},
  eprinttype = {jstor},
  pages = {383--417},
  publisher = {[American Finance Association, Wiley]},
  issn = {0022-1082},
  doi = {10.2307/2325486},
  urldate = {2026-01-23},
  jstor = {2325486}
}

@article{fanImpactShanghaiHong2017,
  title = {The Impact of {{Shanghai}}--{{Hong Kong Stock Connect}} Policy on {{A-H}} Share Price Premium},
  author = {Fan, Qingliang and Wang, Ting},
  year = 2017,
  month = may,
  journal = {Finance Research Letters},
  volume = {21},
  pages = {222--227},
  issn = {15446123},
  doi = {10.1016/j.frl.2016.11.014},
  urldate = {2026-01-04},
  language = {en}
}

@article{gagnonMultimarketTradingArbitrage2010,
  title = {Multi-Market Trading and Arbitrage},
  author = {Gagnon, Louis and Karolyi, G.Andrew},
  year = 2010,
  month = jul,
  journal = {Journal of Financial Economics},
  volume = {97},
  number = {1},
  pages = {53--80},
  issn = {0304405X},
  doi = {10.1016/j.jfineco.2010.03.005},
  urldate = {2026-01-23},
  copyright = {https://www.elsevier.com/tdm/userlicense/1.0/},
  language = {en}
}

@article{mengEffectOverseasInvestors2023,
  title = {The Effect of Overseas Investors on Local Market Efficiency: Evidence from the {{Shanghai}}/{{Shenzhen}}--{{Hong Kong Stock Connect}}},
  shorttitle = {The Effect of Overseas Investors on Local Market Efficiency},
  author = {Meng, Yan and Xiong, Lingyun and Xiao, Lijuan and Bai, Min},
  year = 2023,
  month = jan,
  journal = {Financial Innovation},
  volume = {9},
  number = {1},
  pages = {42},
  issn = {2199-4730},
  doi = {10.1186/s40854-022-00429-3},
  urldate = {2026-01-09},
  language = {en}
}

@article{mittonStockMarketLiberalization2006,
  title = {Stock Market Liberalization and Operating Performance at the Firm Level},
  author = {Mitton, Todd},
  year = 2006,
  month = sep,
  journal = {Journal of Financial Economics},
  volume = {81},
  number = {3},
  pages = {625--647},
  issn = {0304-405X},
  doi = {10.1016/j.jfineco.2005.09.001},
  urldate = {2026-02-04}
}

@article{zhangMarketReactionCrossborder2022,
  title = {The {{Market Reaction}} to {{Cross}}-border {{Listings}}: {{Evidence}} from {{AH Listed Firms}}},
  shorttitle = {The {{Market Reaction}} to {{Cross}}-border {{Listings}}},
  author = {Zhang, John Fan},
  year = 2022,
  month = nov,
  journal = {China \& World Economy},
  volume = {30},
  number = {6},
  pages = {183--218},
  issn = {1671-2234, 1749-124X},
  doi = {10.1111/cwe.12451},
  urldate = {2026-04-14},
  language = {en}
}

\newpage
\begin{sidewaystable}[htbp]
\centering
\caption{Firm-Level Characteristics of the Sample}
\label{tab:1}
\begin{threeparttable}
\resizebox{\linewidth}{!}{%
\begin{tabular}{lcccc}
\toprule
\multicolumn{5}{l}{\textit{Panel A. Industry composition}} \\
\midrule
Industry & \multicolumn{2}{c}{Number of firms} & \multicolumn{2}{c}{Percentage (\%)} \\
\midrule
Industrials & \multicolumn{2}{c}{22} & \multicolumn{2}{c}{32.84} \\
Financials & \multicolumn{2}{c}{15} & \multicolumn{2}{c}{22.39} \\
Materials & \multicolumn{2}{c}{9} & \multicolumn{2}{c}{13.43} \\
Energy & \multicolumn{2}{c}{9} & \multicolumn{2}{c}{13.43} \\
Utilities & \multicolumn{2}{c}{4} & \multicolumn{2}{c}{5.97} \\
Health Care & \multicolumn{2}{c}{3} & \multicolumn{2}{c}{4.48} \\
Consumer Discretionary & \multicolumn{2}{c}{2} & \multicolumn{2}{c}{2.99} \\
Consumer Staples & \multicolumn{2}{c}{1} & \multicolumn{2}{c}{1.49} \\
Communication Services & \multicolumn{2}{c}{1} & \multicolumn{2}{c}{1.49} \\
Real Estate & \multicolumn{2}{c}{1} & \multicolumn{2}{c}{1.49} \\
\midrule
Total & \multicolumn{2}{c}{67} & \multicolumn{2}{c}{100.00} \\
\midrule
\multicolumn{5}{l}{\textit{Panel B. Listing-path characteristics}} \\
\midrule
Listing path & Number of firms & Mean interval (days) & Mean log A-market cap (CNY)& Mean log H-market cap (HKD) \\
\midrule
A$\rightarrow$H & 14 & 2071.79 & 23.87 & 23.47 \\
H$\rightarrow$A & 51 & 1408.41 & 24.08 & 21.96 \\
Simultaneous & 2 & 0.00 & 26.78 & 24.88 \\
\midrule
Total & 67 &  &  &  \\
\bottomrule
\end{tabular}%
}
\begin{tablenotes}
\footnotesize
\item Note: This table reports firm-level characteristics for the 67 A-H dual-listed firms in our sample. Panel A presents the industry distribution based on GICS classifications. Panel B presents firms' listing path, including the number of firms in each group, the average interval between A-share and H-share listing dates, and the average log market capitalization at the time of listing in the A-share and H-share markets. The sample is concentrated in Industrials and Financials, and most firms first listed in H share market and dual-listed back to A share market (H-to-A firms). Compared to H-to-A firms, A-to-H firms exhibit a longer average interval between the two listings.
\end{tablenotes}
\end{threeparttable}
\label{tab:firm_characteristics}
\end{sidewaystable}

\newpage
\begin{table}[htbp]
\centering
\caption{Descriptive Statistics}
\label{tab:2}
\setlength{\tabcolsep}{13pt}
\renewcommand{\arraystretch}{1.2} 
\begin{threeparttable}
\begin{tabular}{lcccccc}
\toprule
Variable & Mean & Max & Min & Median & SD & N \\
\midrule
\texttt{Prem}          & 0.683 & 7.759   & -0.346  & 0.462  & 0.832 & 6,462 \\
\texttt{SHHKPolicy}   & 0.558 & 1.000   & 0.000   & 1.000  & 0.497 & 6,462 \\
\texttt{Effboth}       & -0.015 & -0.002 & -0.067  & -0.014 & 0.006 & 6,462 \\
\texttt{Turnover}      & 2.009 & 120.209 & 0.012   & 1.035  & 3.429 & 6,462 \\
\texttt{SOE}           & 0.919 & 1.000   & 0.000   & 1.000  & 0.273 & 6,462 \\
\texttt{Demand}        & 2.928 & 48.046  & 0.040   & 2.614  & 1.951 & 6,462 \\
\texttt{Size}          & 25.028 & 28.555 & 21.262  & 24.906 & 1.456 & 6,462 \\
\texttt{Interest rate}  & 9.020 & 23.380  & 1.355   & 9.095  & 5.414 & 6,462 \\
\bottomrule
\end{tabular}

\begin{tablenotes}
\footnotesize
\item \textit{Notes}: This table presents the descriptive statistics for the variables used in this paper, covering a total of 6,462 firm-month observations. \texttt{Prem} is the monthly AH share price premium. \texttt{SHHKPolicy} is a binary indicator equal to one for the period following the implementation of the Shanghai-Hong Kong Stock Connect in November 2014 and zero otherwise. \texttt{Effboth} represents the aggregated market efficiency for the Mainland and Hong Kong markets. \texttt{Turnover} denotes the stock turnover ratio, capturing trading liquidity. \texttt{SOE} is a dummy variable equal to one for state-owned enterprises. \texttt{Demand} refers to the market demand for shares. \texttt{Size} represents the firm size, measured as the natural logarithm of total market capitalization. \texttt{Interest rate} is the prevailing market interest rate. Mean, maximum (Max), minimum (Min), median, and standard deviation (SD) are provided for each variable.
\end{tablenotes}
\end{threeparttable}
\end{table}

\newpage
\begin{sidewaystable}
\centering
\caption{Baseline Effect of the SHHK Stock Connect on the A-H Share Price Premium}
\label{tab:3}
\begin{threeparttable}

{
\def\sym#1{\ifmmode^{#1}\else\(^{#1}\)\fi}

\begin{tabular}{l*{6}{c}}

\toprule
            &\multicolumn{1}{c}{AH Premium}&\multicolumn{1}{c}{AH Premium}&\multicolumn{1}{c}{AH Premium}&\multicolumn{1}{c}{AH Premium}&\multicolumn{1}{c}{AH Premium}&\multicolumn{1}{c}{AH Premium}\\
            \cmidrule(lr){2-7}
            &\multicolumn{1}{c}{(1)}&\multicolumn{1}{c}{(2)}&\multicolumn{1}{c}{(3)}&\multicolumn{1}{c}{(4)}&\multicolumn{1}{c}{(5)}&\multicolumn{1}{c}{(6)}\\
\midrule
\texttt{Lagged Premium}      
& \textbf{0.727\sym{***}} 
& \textbf{0.676\sym{***}} 
& \textbf{0.677\sym{***}} 
& \textbf{0.676\sym{***}} 
& \textbf{0.693\sym{***}} 
& \textbf{0.694\sym{***}} \\
            & (0.130) 
            & (0.144) 
            & (0.142) 
            & (0.142) 
            & (0.136) 
            & (0.136) \\
\addlinespace

\texttt{SHHKPolicy}  
& \textbf{0.184\sym{***}}
& \textbf{0.145\sym{***}}
& \textbf{0.144\sym{***}}
& \textbf{0.147\sym{***}}
& \textbf{0.180\sym{***}}
& \textbf{0.184\sym{***}} \\
            & (0.017) 
            & (0.025) 
            & (0.024) 
            & (0.023) 
            & (0.022) 
            & (0.022) \\
\addlinespace

\texttt{Turnover}    
&                     
& 0.025               
& \textbf{0.025\sym{*}}  
& \textbf{0.024\sym{*}}  
& \textbf{0.017\sym{**}} 
& \textbf{0.017\sym{**}} \\
            &                     
            & (0.015) 
            & (0.015) 
            & (0.014) 
            & (0.009) 
            & (0.009) \\
\addlinespace

\texttt{SOE}         
&                     
&                     
& 0.055         
& 0.053         
& 0.030         
& 0.030         \\
            &                     
            &                     
            & (0.059)         
            & (0.062)         
            & (0.050)         
            & (0.049)         \\
\addlinespace

\texttt{Demand}      
&                     
&                     
&                     
& -0.012         
& -0.005         
& -0.005         \\
            &                     
            &                     
            &                     
            & (0.014)         
            & (0.012)         
            & (0.012)         \\
\addlinespace

\texttt{Size}        
&                     
&                     
&                     
&                     
& \textbf{-0.075\sym{*}}  
& \textbf{-0.075\sym{*}}  \\
            &                     
            &                     
            &                     
            &                     
            & (0.044)         
            & (0.043)         \\
\addlinespace

\texttt{Interest rate}
&                     
&                     
&                     
&                     
&                     
& 0.002         \\
            &                     
            &                     
            &                     
            &                     
            &                     
            & (0.002)         \\
\midrule

Observations&    6363         &    6363        &    6363         &    6363         &    6363         &    6363        \\
AR(2) $p$-value &       0.315         &       0.912         &       0.915         &       0.891         &       0.797         &       0.758         \\
Hansen Test $p$-value &       0.548         &       0.416         &       0.417         &       0.425         &       0.462         &       0.469         \\
\bottomrule
\end{tabular}
}
\begin{tablenotes}
\footnotesize
\item \textit{Notes}: This table reports the two-step system GMM estimation results investigating the baseline effect of the Shanghai-Hong Kong (SHHK) Stock Connect on the AH price premium. Columns (1)–(6) gradually incorporate firm-level control variables. The dependent variable is the monthly AH price premium. \texttt{SHHKPolicy} is a binary indicator equal to one for the period following the implementation of the SHHK Stock Connect in November 2014 and zero otherwise. \texttt{Turnover}, \texttt{SOE}, \texttt{Demand}, \texttt{Size}, and \texttt{Interest rate} are included as control variables. Endogenous variables are instrumented using their own lags dated $t-2$ and $t-3$. The $p$-values for the AR(2) and Hansen tests are provided to verify the absence of second-order serial correlation and the validity of the instruments, respectively. Firm and year fixed effects are included. Robust standard errors are reported in parentheses.
\item $^{***}$, $^{**}$, and $^{*}$ denote statistical significance at the 1\%, 5\%, and 10\% levels, respectively.

\end{tablenotes}
\end{threeparttable}
\end{sidewaystable}

\begin{sidewaystable}[htbp]
\centering
\caption{Efficiency-Dependent Effects of the SHHK Stock Connect}
\label{tab:4}
\setlength{\tabcolsep}{100pt}
   \renewcommand{\arraystretch}{1.1}
\begin{threeparttable}
\begin{tabular}{l c}
\toprule
 & \multicolumn{1}{c}{AH Premium} \\
\cmidrule(lr){2-2}
 & (1) \\
\midrule
\texttt{Lagged Premium} 
    & \textbf{0.633\sym{***}} \\
    & (0.081) \\

\texttt{SHHKPolicy} 
    & \textbf{-0.490\sym{**}} \\
    & (0.196) \\

\texttt{Effboth} 
    & \textbf{39.554\sym{**}} \\
    & (15.105) \\

\texttt{SHHKPolicy} $\times$ \texttt{Effboth} 
    & \textbf{-45.900\sym{***}} \\
    & (14.584) \\

\texttt{Turnover} 
    & \textbf{0.022\sym{***}} \\
    & (0.007) \\

\texttt{SOE} 
    & 0.021 \\
    & (0.045) \\

\texttt{Demand} 
    & -0.009 \\
    & (0.011) \\

\texttt{Firm Size} 
    & \textbf{-0.108\sym{***}} \\
    & (0.031) \\

\texttt{Interest Rate} 
    & \textbf{0.007\sym{***}} \\
    & (0.002) \\
\midrule

Observations                        & 6,363    \\
AR(2) $p$-value                     & 0.211    \\
Hansen Test $p$-value               & 0.780    \\

\bottomrule
\end{tabular}

\begin{tablenotes}
\footnotesize
\item \textit{Notes}: This table reports the two-step system GMM estimation results investigating how market efficiency moderates the impact of the Shanghai-Hong Kong (SHHK) Stock Connect on the AH price premium. The dependent variable is the monthly AH price premium. \texttt{SHHKPolicy} is a binary indicator equal to one for the period following the implementation of the SHHK Stock Connect in November 2014 and zero otherwise. \texttt{Effboth} represents the aggregated market efficiency for the Mainland and Hong Kong markets. The interaction term \texttt{SHHKPolicy} $\times$ \texttt{Effboth} captures the efficiency-dependent effects of the policy. \texttt{Turnover}, \texttt{SOE}, \texttt{Demand}, \texttt{Size}, and \texttt{Interest Rate} are included as control variables. Endogenous variables are instrumented using their own lags dated $t-2$ and $t-3$. The $p$-values for the AR(2) and Hansen tests are provided to verify the absence of second-order serial correlation and the validity of the instruments, respectively. Firm and year fixed effects are included. Robust standard errors are reported in parentheses.
\item $^{***}$, $^{**}$, and $^{*}$ denote statistical significance at the 1\%, 5\%, and 10\% levels, respectively.
\end{tablenotes}
\end{threeparttable}
\end{sidewaystable}
\clearpage

\newpage
\appendix
\renewcommand{\thesection}{\Alph{section}} 
\counterwithin{table}{section} 
\renewcommand{\thetable}{\thesection\arabic{table}}

\section{Firm list and IPO time line}\label{app:A}

\begin{sidewaysfigure}[p]
    \centering
    \begin{minipage}{0.45\linewidth}
        \centering
        \includegraphics[width=\linewidth]{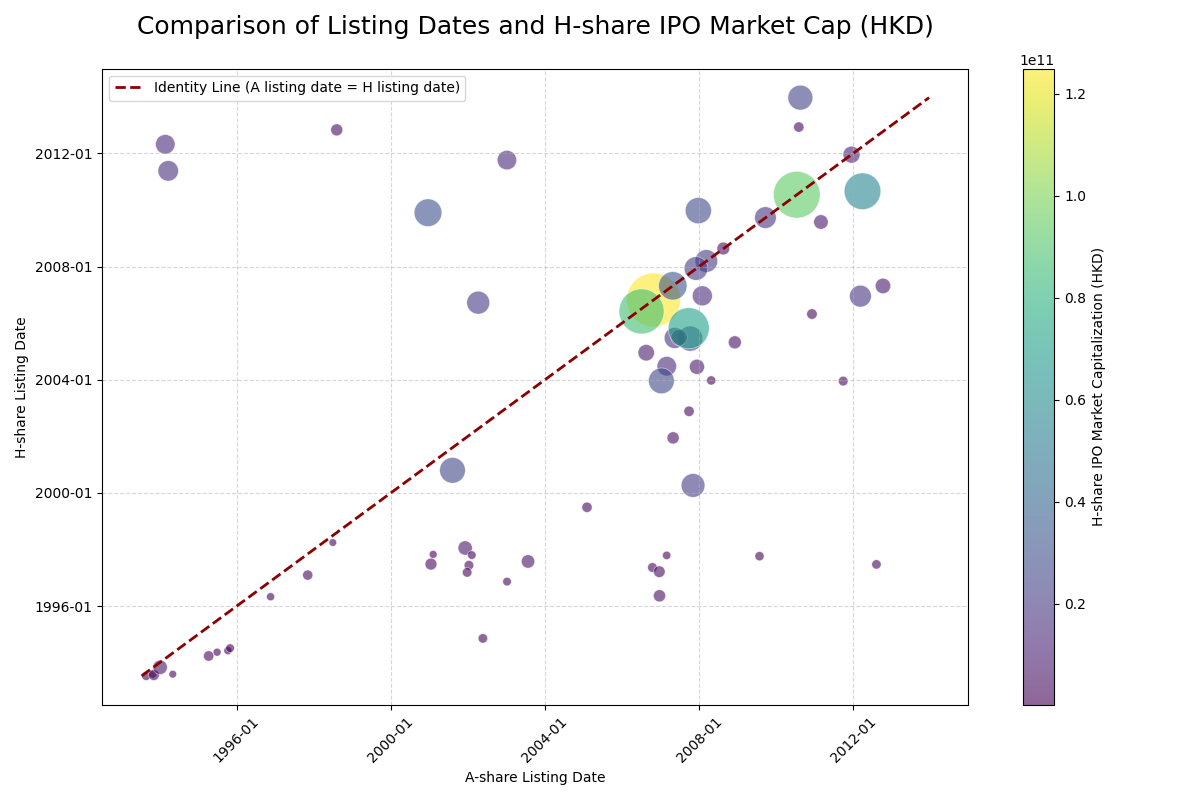}
        \caption{Timeline of the dual-listing process. Bubble size and color represent market capitalization at the time of Hong Kong listing, measured in HKD.}
        \label{fig:Scatter_HKD}
    \end{minipage}
    \hspace{0.05\linewidth}
    \begin{minipage}{0.45\linewidth}
        \centering
        \includegraphics[width=\linewidth]{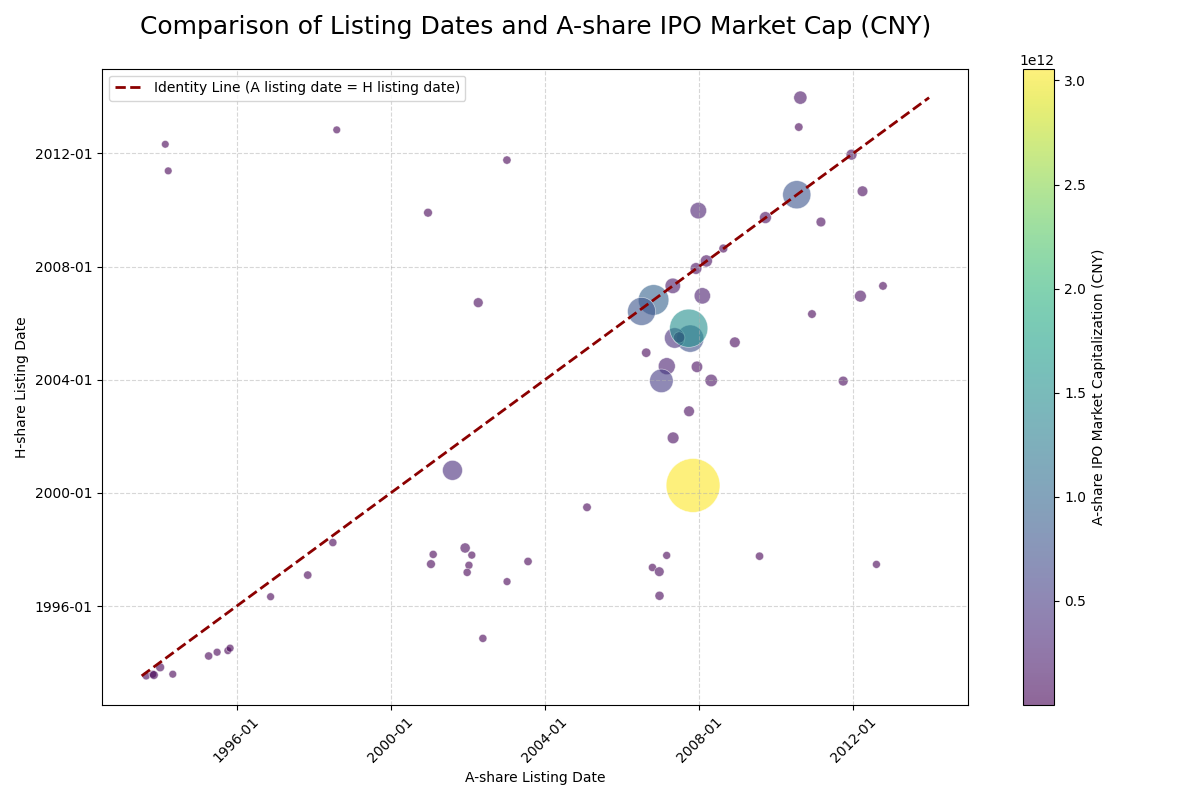}
        \caption{Timeline of the dual-listing process. Bubble size and color represent market capitalization at the time of Shanghai listing, measured in CNY.}
        \label{fig:Scatter_CNY}
    \end{minipage}
\end{sidewaysfigure}

\begin{sidewaysfigure}[p]  
    \centering
    \includegraphics[width=0.7\linewidth]{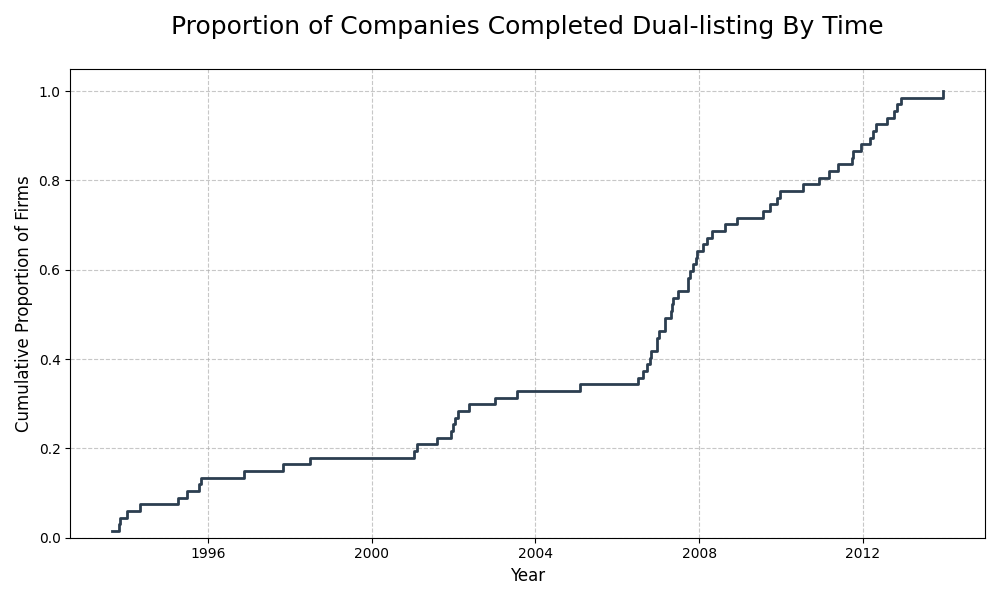}
    \caption{The cumulated proportion of firms that completed the A-H dual listing by time}
    \label{fig:CDF}
\end{sidewaysfigure}

Figure~\ref{fig:Scatter_HKD} and~\ref{fig:Scatter_CNY} illustrate the dual-listing timeline for the 67 firms in our sample. In both figures, the x-axis and y-axis represent the A-share and H-share listing dates, respectively. The bubble size corresponds to the firm’s market capitalization at the time of its IPO. Specifically, Figure~\ref{fig:Scatter_HKD} scales the bubbles based on market capitalization denominated in Hong Kong Dollars (HKD) at the time of listing in Hong Kong, whereas Figure~\ref{fig:Scatter_CNY} utilizes market capitalization in Chinese Yuan (CNY) at the time of listing in Shanghai.
From these two figures, it is easy to notice that a significant portion of firms completed their listings between 2004 and 2012, with a notable clustering around 2008. Furthermore, majority of bubbles (firms) are positioned below the 45-degree identity line, indicating these firms finished A-H dual listing by H-to-A listing path. To be more specific, these firms initially listed in Hong Kong then dual-listed back to Shanghai. Interestingly, firms with larger market capitalizations exhibit a smaller time gap between listings, as evidenced by their closer proximity to the identity line. To better understand the trend of A-H dual listing in our sample, we construct Cumulative Distribution Function (CDF) of the completion time of dual-listing as shown in Figure~\ref{fig:CDF}. The earliest A-H dual-listed firm is in 1994, gradually, over 30 percent of firms finished A-H dual listing before 2006. After 2006, more and more firms finished A-H dual listing, especially in the year of 2007. 

Figure~\ref{fig:CDF} presents the cumulative distribution function (CDF) of dual-listing completion time for firms in our sample. The earliest A–H dual listing occurred in 1994. The distribution suggests that dual listings accumulated gradually in the early period, with slightly more than 30\% of sample firms having completed the process by 2006. After 2006, the pace accelerated, with a noticeable increase in completions around 2007.

\newpage
{\footnotesize
\begin{longtable}{lllll}
\caption{A-H Dual-Listed Firms}
\label{tab:ah_dual_listing_long} \\
\hline
A-share Code & H-share Code & Firm Name & A-listing Date & H-listing Date \\
\hline
\endfirsthead

\multicolumn{5}{l}{\textit{Table \ref{tab:ah_dual_listing_long} continued}} \\
\hline
A-share Code & H-share Code & Firm Name & A-listing Date & H-listing Date \\
\hline
\endhead

\hline
\endfoot

\endlastfoot

600011.SH & 0902.HK & HPI & 2001-12-06 & 1998-01-21 \\
600012.SH & 0995.HK & ANHUI EXPRESSWAY & 2003-01-07 & 1996-11-13 \\
600016.SH & 1988.HK & CMBC & 2000-12-19 & 2009-11-26 \\
600026.SH & 1138.HK & COSCO SHIPPING Energy & 2002-05-23 & 1994-11-11 \\
600027.SH & 1071.HK & HDPI & 2005-02-03 & 1999-06-30 \\
600028.SH & 0386.HK & Sinopec Corp. & 2001-08-08 & 2000-10-19 \\
600029.SH & 1055.HK & CSN & 2003-07-25 & 1997-07-31 \\
600030.SH & 6030.HK & CITIC Securities Co., Ltd. & 2003-01-06 & 2011-10-06 \\
600036.SH & 3968.HK & CM Bank & 2002-04-09 & 2006-09-22 \\
600115.SH & 0670.HK & CEA & 1997-11-05 & 1997-02-05 \\
600188.SH & 1171.HK & Yankuang Energy & 1998-07-01 & 1998-04-01 \\
600196.SH & 2196.HK & Fosun Pharma & 1998-08-07 & 2012-10-30 \\
600332.SH & 0874.HK & GYBYS & 2001-02-06 & 1997-10-30 \\
600362.SH & 0358.HK & JCCL & 2002-01-11 & 1997-06-12 \\
600377.SH & 0177.HK & Jiangsu Expressway & 2001-01-16 & 1997-06-27 \\
600548.SH & 0548.HK & SZEW & 2001-12-25 & 1997-03-12 \\
600585.SH & 0914.HK & ACC & 2002-02-07 & 1997-10-21 \\
600600.SH & 0168.HK & Tsingtao Brewery & 1993-08-27 & 1993-07-15 \\
600685.SH & 0317.HK & COMEC & 1993-10-28 & 1993-08-06 \\
600688.SH & 0338.HK & SPC & 1993-11-08 & 1993-07-26 \\
600775.SH & 0553.HK & NPEC & 1996-11-18 & 1996-05-02 \\
600808.SH & 0323.HK & MAS C.L. & 1994-01-06 & 1993-11-03 \\
600837.SH & 6837.HK & Haitong Securities & 1994-02-24 & 2012-04-27 \\
600860.SH & 0187.HK & Jingcheng Machinery & 1994-05-06 & 1993-08-06 \\
600871.SH & 1033.HK & SSC & 1995-04-11 & 1994-03-29 \\
600874.SH & 1065.HK & TCEPC & 1995-06-30 & 1994-05-17 \\
600875.SH & 1072.HK & DEC & 1995-10-10 & 1994-06-06 \\
600876.SH & 1108.HK & Triumph New Energy & 1995-10-31 & 1994-07-08 \\
601005.SH & 1053.HK & CISC & 2007-02-28 & 1997-10-17 \\
601038.SH & 0038.HK & First Tractor & 2012-08-08 & 1997-06-23 \\
601088.SH & 1088.HK & China Shenhua & 2007-10-09 & 2005-06-15 \\
601107.SH & 0107.HK & Sichuan Expressway & 2009-07-27 & 1997-10-07 \\
601111.SH & 0753.HK & Air China & 2006-08-18 & 2004-12-15 \\
601186.SH & 1186.HK & CRCC & 2008-03-10 & 2008-03-13 \\
601238.SH & 2238.HK & GAC Group & 2012-03-29 & 2010-08-30 \\
601288.SH & 1288.HK & Agricultural Bank of China & 2010-07-15 & 2010-07-16 \\
601318.SH & 2318.HK & Ping An & 2007-03-01 & 2004-06-24 \\
601328.SH & 3328.HK & Bank of Communications & 2007-05-15 & 2005-06-23 \\
601333.SH & 0525.HK & Guangshen Railway & 2006-12-22 & 1996-05-14 \\
601336.SH & 1336.HK & NCI & 2011-12-16 & 2011-12-15 \\
601390.SH & 0390.HK & China Railway & 2007-12-03 & 2007-12-07 \\
601398.SH & 1398.HK & ICBC & 2006-10-27 & 2006-10-27 \\
601588.SH & 0588.HK & Beijing North Star & 2006-10-16 & 1997-05-14 \\
601600.SH & 2600.HK & Chalco & 2007-04-30 & 2001-12-12 \\
601601.SH & 2601.HK & CPIC & 2007-12-25 & 2009-12-23 \\
601607.SH & 2607.HK & Shanghai Pharma & 1994-03-24 & 2011-05-20 \\
601618.SH & 1618.HK & MCC & 2009-09-21 & 2009-09-24 \\
601628.SH & 2628.HK & China Life & 2007-01-09 & 2003-12-18 \\
601633.SH & 2333.HK & Great Wall Motor & 2011-09-28 & 2003-12-15 \\
601717.SH & 0564.HK & ZCZL & 2010-08-03 & 2012-12-05 \\
601727.SH & 2727.HK & Shanghai Electric & 2008-12-05 & 2005-04-28 \\
601766.SH & 1766.HK & CRRC & 2008-08-18 & 2008-08-21 \\
601800.SH & 1800.HK & CCCC & 2012-03-09 & 2006-12-15 \\
601808.SH & 2883.HK & COSL & 2007-09-28 & 2002-11-20 \\
601818.SH & 6818.HK & CEB Bank & 2010-08-18 & 2013-12-20 \\
601857.SH & 0857.HK & PetroChina & 2007-11-05 & 2000-04-07 \\
601866.SH & 2866.HK & COSCO Shipping Development & 2007-12-12 & 2004-06-16 \\
601880.SH & 2880.HK & Liaoning Port & 2010-12-06 & 2006-04-28 \\
601898.SH & 1898.HK & China Coal Energy & 2008-02-01 & 2006-12-19 \\
601899.SH & 2899.HK & Zijin Mining & 2008-04-25 & 2003-12-23 \\
601919.SH & 1919.HK & COSCO Shipping Holdings & 2007-06-26 & 2005-06-30 \\
601939.SH & 0939.HK & CCB & 2007-09-25 & 2005-10-27 \\
601988.SH & 3988.HK & Bank of China & 2006-07-05 & 2006-06-01 \\
601991.SH & 0991.HK & Datang Power & 2006-12-20 & 1997-03-21 \\
601992.SH & 2009.HK & BBMG & 2011-03-01 & 2009-07-29 \\
601998.SH & 0998.HK & CNCB & 2007-04-27 & 2007-04-27 \\
603993.SH & 3993.HK & CMOC & 2012-10-09 & 2007-04-26 \\

\hline
\multicolumn{5}{p{\dimexpr\textwidth-2\tabcolsep\relax}}{\textit{Notes:} This table lists the 67 A-H dual-listed firms included in our sample, together with their A-share and H-share stock codes and listing dates in the Shanghai and Hong Kong markets, respectively. Firms are included only if they completed their dual listing \textit{before} the implementation of the SHHK Stock Connect.} \\

\end{longtable}
}

\newpage
\section{Corwin and Schultz Bid-Ask Spread Estimation}
\label{app:B}
Following the literature, we construct following equations to compute daily bid-ask spread.

\begin{equation}\label{eq:1}
\mathbb{E}\left\{
\sum_{j=0}^{1}
\left[
\ln\!\left(
\frac{\text{H}^{0}_{t+j}}{\text{L}^{0}_{t+j}}
\right)
\right]^2
\right\}
=
2k_1 \sigma_{\text{HL}}^2
+ 4k_2 \sigma_{\text{HL}}
\ln\!\left(\frac{2+\text{S}}{2-\text{S}}\right)
+ 2\left[
\ln\!\left(\frac{2+\text{S}}{2-\text{S}}\right)
\right]^2 ,
\end{equation}

and

\begin{equation}\label{eq:2}
\mathbb{E}\left\{
\left[
\ln\!\left(
\frac{\text{H}^{0}_{t,t+1}}{\text{L}^{0}_{t,t+1}}
\right)
\right]^2
\right\}
=
2k_1 \sigma_{\text{HL}}^2
+ 2\sqrt{2}k_2 \sigma_{\text{HL}}
\ln\!\left(\frac{2+\text{S}}{2-\text{S}}\right)
+ \left[
\ln\!\left(\frac{2+\text{S}}{2-\text{S}}\right)
\right]^2 ,
\end{equation}
where $\text{S}$ denotes the bid-ask spread; $\text{H}^{0}_{t+j}$ and $\text{L}\textbf{}^{0}_{t+j}$ are the
daily highest and lowest prices of a stock at the trading day of $t+j$ ,
respectively; $\text{H}^{0}_{t,t+1}$ and $\text{L}^{0}_{t,t+1}$ are the daily highest and lowest
prices of a stock observed on two consecutive trading days $t$ and $t+1$, respectively;
$\sigma_{HL}$ denotes volatility and $k_1$ and $k_2$ are coefficients.

It is not hard to see that Equation~\ref{eq:1} is a one-day range of the spread ($\text{S}$) at the time of $t+j$. Equation~\ref{eq:2} is a two-day range of the spread ($\text{S}$) from the time of $t$ to $t+1$.

Additionally, in order to make the process easier and in line with \citet{corwinSimpleWayEstimate2012}, we define following:
\begin{equation}\label{eq:3}
\beta
=
\mathbb{E}\left\{
\sum_{j=0}^{1}
\left[
\ln\!\left(
\frac{H^{0}_{t+j}}{L^{0}_{t+j}}
\right)
\right]^2
\right\},
\qquad
\gamma
=
\mathbb{E}\left\{
\left[
\ln\!\left(
\frac{H^{0}_{t,t+1}}{L^{0}_{t,t+1}}
\right)
\right]^2
\right\}
\end{equation}
Solving Equations~\ref{eq:1} and Equation~\ref{eq:2}, we can obtain the spread ($\text{S}$) equation, which is our closed form equation.
\begin{equation*}
S = \frac{2\left(e^{\alpha}-1\right)}{1+e^{\alpha}}, 
\end{equation*}
where
\begin{equation*}
\alpha = \frac{\sqrt{2\beta}-\sqrt{\beta}}{3-2\sqrt{2}} - \sqrt{\frac{\gamma}{3-2\sqrt{2}}}.
\end{equation*}
After computing daily bid-ask spreads, we aggregate them to the monthly level by taking the monthly average to align with the frequency of the other variables.\footnote{For negative daily bid–ask spread estimates ($\text{S}$), we follow \citet{corwinSimpleWayEstimate2012} and set negative values to zero prior to calculating monthly averages. This approach yields more accurate estimates than either retaining negative spread observations or excluding them from the sample.} In line with \citet{mengEffectOverseasInvestors2023}, we use the negative of the bid-ask spread to proxy for market efficiency, implying an inverse relationship between the bid-ask spread and market efficiency. Additionally, we construct an aggregate efficiency measure, $\texttt{Effboth}$, by summing the A-share and H-share market efficiency proxies to capture the overall information environment of A-H dual-listed firms.\footnote{It is worth noting that $Effboth$ is strictly negative in our sample by construction. Values closer to zero indicate higher market efficiency, while more negative values correspond to lower efficiency.}
After computing daily bid-ask spreads, we aggregate them to the monthly level by taking the monthly average to align with the frequency of the other variables. 


\newpage
\section{Announcement Effect to the share price premium}
\label{app:C}
Table~\ref{tab:B1} reports GMM estimates across specifications in Columns~(1)-(6), where control variables are sequentially added and Column~(6) presents the full model. Across all columns, the estimated coefficient on \texttt{Announcement} is close to zero and statistically insignificant, suggesting no economically meaningful change in the AH premium around the announcement date at the monthly frequency. A potential interpretation is that the announcement did not immediately relax the trading, settlement, or capital-account constraints that limit cross-market arbitrage; consequently, any convergence forces induced by the policy were unlikely to materialize until implementation. 
\begin{sidewaystable}[htbp]
\centering
\caption{Announcement Effects of the SHHK Stock Connect on the A–H Premium}
\label{tab:B1}
\begin{threeparttable}
{
\def\sym#1{\ifmmode^{#1}\else\(^{#1}\)\fi}
\begin{tabular}{l*{6}{c}}
\toprule
            &\multicolumn{1}{c}{AH Premium}&\multicolumn{1}{c}{AH Premium}&\multicolumn{1}{c}{AH Premium}&\multicolumn{1}{c}{AH Premium}&\multicolumn{1}{c}{AH Premium}&\multicolumn{1}{c}{AH Premium}\\
            \cmidrule(lr){2-7}
            &\multicolumn{1}{c}{(1)}&\multicolumn{1}{c}{(2)}&\multicolumn{1}{c}{(3)}&\multicolumn{1}{c}{(4)}&\multicolumn{1}{c}{(5)}&\multicolumn{1}{c}{(6)}\\
\midrule
\texttt{Lagged Premium}       
& \textbf{0.788\sym{***}}
& \textbf{0.725\sym{***}}
& \textbf{0.726\sym{***}}
& \textbf{0.726\sym{***}}
& \textbf{0.748\sym{***}}
& \textbf{0.748\sym{***}} \\
            & (0.115)
            & (0.125)
            & (0.124)
            & (0.126)
            & (0.115)
            & (0.115) \\
\addlinespace

\texttt{Announcement}
& 0.001
& -0.008
& -0.008
& -0.007
& 0.004
& 0.004 \\
            & (0.019)
            & (0.022)
            & (0.022)
            & (0.021)
            & (0.015)
            & (0.017) \\
\addlinespace

\texttt{Turnover}    
&                     
& \textbf{0.023\sym{*}}
& \textbf{0.023\sym{*}}
& \textbf{0.022\sym{*}}
& \textbf{0.016\sym{**}}
& \textbf{0.016\sym{**}} \\
            &                     
            & (0.013)
            & (0.013)
            & (0.012)
            & (0.007)
            & (0.007) \\
\addlinespace

\texttt{SOE}         
&                     
&                     
& 0.048
& 0.046
& 0.027
& 0.027 \\
            &                     
            &                     
            & (0.048)
            & (0.051)
            & (0.039)
            & (0.039) \\
\addlinespace

\texttt{Demand}     
&                     
&                     
&                     
& -0.009
& -0.004
& -0.004 \\
            &                     
            &                     
            &                     
            & (0.012)
            & (0.009)
            & (0.009) \\
\addlinespace

\texttt{Size}       
&                     
&                     
&                     
&                     
& \textbf{-0.060\sym{*}}
& \textbf{-0.060\sym{*}} \\
            &                     
            &                     
            &                     
            &                     
            & (0.036)
            & (0.036) \\
\addlinespace

\texttt{Interest rate}
&                     
&                     
&                     
&                     
&                     
& -0.000 \\
            &                     
            &                     
            &                     
            &                     
            &                     
            & (0.002) \\
\midrule

Observations&    6363         &    6363        &    6363         &    6363         &    6363         &    6363        \\
AR(2) $p$-value &       0.152         &       0.621         &       0.627         &       0.596         &       0.449         &       0.452         \\
Hansen Test $p$-value &       0.522         &       0.404         &       0.405         &       0.413         &       0.442         &       0.441         \\
\bottomrule
\end{tabular}
}

\begin{tablenotes}
\footnotesize
\item \textit{Notes}: This table reports the two-step system GMM estimation results investigating the impact of the SHHK Stock Connect announcement on the AH price premium. Columns (1)–(6) gradually incorporate firm-level control variables to assess the robustness of the baseline specification. The dependent variable is the monthly AH price premium. \texttt{Announcement} is a binary indicator equal to one for the period of the SHHK Stock Connect announcement (after April 2014) and zero otherwise. \texttt{Turnover}, \texttt{SOE}, \texttt{Demand}, \texttt{Size}, and \texttt{Interest rate} are included as control variables. Endogenous variables are instrumented using their own lags dated $t-2$ and $t-3$. The $p$-values for the AR(2) and Hansen tests are provided to verify the absence of second-order serial correlation and the validity of the instruments, respectively. Firm and year fixed effects are included. Robust standard errors are reported in parentheses.
\item $^{***}$, $^{**}$, and $^{*}$ denote statistical significance at the 1\%, 5\%, and 10\% levels, respectively.
\end{tablenotes}
\end{threeparttable}
\end{sidewaystable}

\newpage
\section{Efficiency-Dependent Announcement Effects on the A–H Share Price Premium}
\label{app:D}
Table~\ref{tab:D1} examines whether the announcement of the SHHK Stock Connect has an efficiency-dependent effect on the A-H price premium. The coefficient on \texttt{Announcement} is negative but statistically insignificant ($-0.217$), suggesting no significant change in the AH premium around the announcement date. More importantly, the interaction term \texttt{Announcement} $\times$ \texttt{Effboth} is also statistically insignificant ($-14.471$), providing no evidence that the announcement effect varies systematically with market efficiency. Consistent with these results, the main effect of \texttt{Effboth} is not statistically different from zero in this specification. Overall, Table~\ref{tab:D1} indicates that the announcement itself and the interaction with market efficiency are unlikely to generate a measurable impact on the A-H premium.

\def\sym#1{\ifmmode^{#1}\else\(^{#1}\)\fi}
\begin{sidewaystable}[htbp]
\centering
\caption{Efficiency-Dependent Effects of the Announcement of SHHK Stock Connect}
\label{tab:D1}

\setlength{\tabcolsep}{100pt}
\begin{threeparttable}
\begin{tabular}{l c}
\toprule
 & \multicolumn{1}{c}{AH Premium} \\
\cmidrule(lr){2-2}
 & (1) \\
\midrule
\texttt{Lagged Premium}      
    & \textbf{0.678\sym{***}} \\
    & (0.092) \\
\addlinespace

\texttt{Announcement}
    & -0.217 \\
    & (0.203) \\
\addlinespace

\texttt{Effboth}     
    &  13.023 \\
    & (13.283) \\
\addlinespace

\texttt{Announcement} $\times$ \texttt{Effboth} 
    & -14.471 \\
    & (13.547) \\
\addlinespace

\texttt{Turnover}    
    & \textbf{0.021\sym{***}} \\
    & (0.006) \\
\addlinespace

\texttt{SOE}         
    & 0.026 \\
    & (0.039) \\
\addlinespace

\texttt{Demand}      
    & -0.005 \\
    & (0.009) \\
\addlinespace

\texttt{Size}        
    & \textbf{-0.087\sym{**}} \\
    & (0.033) \\
\addlinespace

\texttt{Interest rate}
    & 0.001 \\
    & (0.002) \\
\midrule

Observations                        & 6,363    \\
AR(2) $p$-value                     & 0.298    \\
Hansen Test $p$-value               & 0.778    \\

\bottomrule
\end{tabular}

\begin{tablenotes}
\footnotesize
\item \textit{Notes}: This table reports the two-step system GMM estimation results investigating how market efficiency moderates the impact of the SHHK Stock Connect announcement on the AH price premium. The dependent variable is the monthly AH price premium. \texttt{Announcement} is a binary indicator equal to one for the period of the SHHK Stock Connect announcement (after April 2014) and zero otherwise.  \texttt{Effboth} represents the aggregated market efficiency for the Mainland and Hong Kong markets. The interaction term \texttt{Announcement} $\times$ \texttt{Effboth} captures the efficiency-dependent effects of the announcement event. \texttt{Turnover}, \texttt{SOE}, \texttt{Demand}, \texttt{Size}, and \texttt{Interest rate} are included as control variables. Endogenous variables are instrumented using their own lags dated $t-2$ and $t-3$. The $p$-values for the AR(2) and Hansen tests are provided to verify the absence of second-order serial correlation and the validity of the instruments, respectively. Firm and year fixed effects are included. Robust standard errors are reported in parentheses.
\item $^{***}$, $^{**}$, and $^{*}$ denote statistical significance at the 1\%, 5\%, and 10\% levels, respectively.
\end{tablenotes}
\end{threeparttable}
\end{sidewaystable}

\newpage
\section{Alternative Market Efficiency Proxy}
\label{app:E}
According to \citet{mengEffectOverseasInvestors2023}, we use an alternative market efficiency proxy, \texttt{Effboth2}, as a robustness check. Specifically, for each trading day $t$ we compute the relative high-low spread as 

\begin{equation}
\text{S}_{2}=\frac{\text{H}_t-\text{L}_t}{\big(\text{H}_t-\text{L}_t\big)/2},
\end{equation}
and then take the month average of this daily measure. Similar to our previous definition regarding market efficiency, we use the negative spread to be a proxy for market efficiency. That is:
\begin{equation}
\text{Eff}_{2}= - \text{S}_{2},
\end{equation}
For A-share market and H-share market, we calculate the $\texttt{Eff}_{2}$ separately and then add them up as an alternative proxy for overall market efficiency ($\texttt{EFFboth2}$). Table~\ref{tab:E1} presents our regression results for third robustness check. 
Even though $\texttt{SHHKPolicy}$ is not statistically significant, the sign of coefficient is negative. The interaction term $\texttt{SHHKPolicy} \times \texttt{Effboth2}$ is negatively significant at 5\% level. So, in general, this result is consistent with the result shown in Table~\ref{tab:3}.   
The SHHK Stock Connect was launched in November, 2014, to further test whether the effect comes from SHHK Stock Connect and assess the robustness of our previous findings, we conduct a placebo test by assigning the implementation of the SHHK Stock Connect to a later date. Specifically, we assume that the policy was introduced two years after its actual launch and construct a binary indicator, $\texttt{FakePolicy}$, which equals one for observations after November 2016 and zero otherwise.

\def\sym#1{\ifmmode^{#1}\else\(^{#1}\)\fi}
\begin{sidewaystable}[htbp]

\centering
\caption{Efficiency-Dependent Effects of the SHHK Stock Connect using alternative efficiency measurement }
\label{tab:E1}
\setlength{\tabcolsep}{100pt}
\begin{threeparttable}
\begin{tabular}{l c}
\toprule
 & \multicolumn{1}{c}{AH Premium} \\
\cmidrule(lr){2-2}
 & (1) \\
\midrule
\texttt{Lagged Premium}     
    & \textbf{0.796\sym{***}} \\
    & (0.066) \\
\addlinespace

\texttt{SHHKPolicy}  
    & -0.130 \\
    & (0.111) \\
\addlinespace

\texttt{Effboth2}        
    & 2.370 \\
    & (1.638) \\
\addlinespace

\texttt{SHHKPolicy} $\times$ \texttt{Effboth2}   
    & \textbf{-4.504\sym{**}} \\
    & (1.859) \\
\addlinespace

\texttt{Turnover}    
    & \textbf{0.011\sym{**}} \\
    & (0.004) \\
\addlinespace

\texttt{SOE}         
    & 0.033 \\
    & (0.031) \\
\addlinespace

\texttt{Demand}      
    & 0.000 \\
    & (0.006) \\
\addlinespace

\texttt{Size}        
    & \textbf{-0.042\sym{**}} \\
    & (0.020) \\
\addlinespace

\texttt{Interest rate}
    & \textbf{0.006\sym{***}} \\
    & (0.001) \\
\midrule

Observations&    6363         \\
AR(2) $p$-value&       0.406         \\
Hansen Test $p$-value&       0.272         \\
\bottomrule
\end{tabular}

\begin{tablenotes}
\footnotesize
\item \textit{Notes}: This table reports the two-step system GMM estimation results investigating how an alternative measure of market efficiency moderates the impact of the Shanghai-Hong Kong (SHHK) Stock Connect on the AH price premium. The dependent variable is the monthly AH price premium. \texttt{SHHKPolicy} is a binary indicator equal to one for the period following the implementation of the SHHK Stock Connect in November 2014 and zero otherwise. \texttt{Effboth2} denotes an alternative measurement of the aggregated pricing efficiency of the Mainland and Hong Kong markets, where higher values indicate greater efficiency. The interaction term \texttt{SHHKPolicy} $\times$ \texttt{Effboth2} captures the efficiency-dependent effects under this alternative specification. \texttt{Turnover}, \texttt{SOE}, \texttt{Demand}, \texttt{Size}, and \texttt{Interest Rate} are included as control variables. Endogenous variables are instrumented using their own lags dated $t-2$ and $t-3$. The $p$-values for the AR(2) and Hansen tests are provided to verify the absence of second-order serial correlation and the validity of the instruments, respectively. Firm and year fixed effects are included. Robust standard errors are reported in parentheses.
\item $^{***}$, $^{**}$, and $^{*}$ denote statistical significance at the 1\%, 5\%, and 10\% levels, respectively.
\end{tablenotes}
\end{threeparttable}
\end{sidewaystable}

\newpage
\section{Placebo Test}\label{app:F}

Table~\ref{tab:F1} reports the placebo regression results for the effect of SHHK Stock Connect on share price premium. In contrast to the baseline estimates, the coefficient on $\texttt{FakePolicy}$ is positive but insignificant across all columns. This finding suggests that share price premium is not driven by common time trend, the SHHK stock policy does affect the premium. Next step, we conducted the placebo test regarding whether there is an efficiency dependent effects of the SHHK Stock Connect. Table~\ref{tab:F2} reports a placebo test in which the SHHK Stock Connect is assumed to have been implemented two years later, in November 2016. The key coefficient of interest is the interaction term between the placebo policy indicator and market efficiency, $\texttt{FakePolicy} \times \texttt{Effboth}$.

The interaction term $\texttt{FakePolicy} \times \texttt{Effboth}$ is statistically insignificant, indicating that when the policy timing is artificially shifted, market efficiency does not systematically condition the relationship between the placebo policy and the AH price premium. This result suggests that the efficiency-dependent effect documented in the previous analysis is not driven by common time trends or spurious correlations, but is instead specific to the actual implementation of the SHHK Stock Connect.

\def\sym#1{\ifmmode^{#1}\else\(^{#1}\)\fi}
\begin{sidewaystable}[htbp]
\centering
\caption{Placebo test for SHHK Stock Connect on A-H premium}
\label{tab:F1}
\begin{threeparttable}
\begin{tabular}{l*{6}{c}}
\toprule
            &\multicolumn{1}{c}{AH Premium}&\multicolumn{1}{c}{AH Premium}&\multicolumn{1}{c}{AH Premium}&\multicolumn{1}{c}{AH Premium}&\multicolumn{1}{c}{AH Premium}&\multicolumn{1}{c}{AH Premium}\\
            \cmidrule(lr){2-7}
            &\multicolumn{1}{c}{(1)}&\multicolumn{1}{c}{(2)}&\multicolumn{1}{c}{(3)}&\multicolumn{1}{c}{(4)}&\multicolumn{1}{c}{(5)}&\multicolumn{1}{c}{(6)}\\
\midrule
\texttt{Lagged Premium}     
& \textbf{0.790\sym{***}}
& \textbf{0.725\sym{***}}
& \textbf{0.726\sym{***}}
& \textbf{0.727\sym{***}}
& \textbf{0.751\sym{***}}
& \textbf{0.751\sym{***}} \\
            & (0.118)
            & (0.126)
            & (0.125)
            & (0.128)
            & (0.117)
            & (0.118) \\
\addlinespace

\texttt{FakePolicy} 
& 0.013
& 0.000
& 0.001
& 0.001
& 0.011
& 0.011 \\
            & (0.030)
            & (0.036)
            & (0.036)
            & (0.036)
            & (0.031)
            & (0.032) \\
\addlinespace

\texttt{Turnover}    
&                     
& \textbf{0.023\sym{*}}
& \textbf{0.023\sym{*}}
& \textbf{0.022\sym{*}}
& \textbf{0.016\sym{**}}
& \textbf{0.016\sym{**}} \\
            &                     
            & (0.013)
            & (0.013)
            & (0.012)
            & (0.007)
            & (0.007) \\
\addlinespace

\texttt{SOE}        
&                     
&                     
& 0.049
& 0.047
& 0.027
& 0.027 \\
            &                     
            &                     
            & (0.047)
            & (0.050)
            & (0.039)
            & (0.038) \\
\addlinespace

\texttt{Demand}      
&                     
&                     
&                     
& -0.009
& -0.004
& -0.004 \\
            &                     
            &                     
            &                     
            & (0.011)
            & (0.009)
            & (0.009) \\
\addlinespace

\texttt{Size}        
&                     
&                     
&                     
&                     
& -0.059
& -0.059 \\
            &                     
            &                     
            &                     
            &                     
            & (0.036)
            & (0.036) \\
\addlinespace

\texttt{Interest rate}
&                     
&                     
&                     
&                     
&                     
& -0.000 \\
            &                     
            &                     
            &                     
            &                     
            &                     
            & (0.002) \\
\midrule

Observations&    6363        &    6363         &    6363         &    6363         &    6363         &    6363         \\
AR(2) $p$-value&       0.158         &       0.615         &       0.621         &       0.589         &       0.450         &       0.452         \\
Hansen Test $p$-value&       0.527         &       0.402         &       0.403         &       0.412         &       0.445         &       0.445         \\\bottomrule
\end{tabular}
\begin{tablenotes}
\footnotesize
\item \textit{Notes}: This table reports the two-step system GMM estimation results for a placebo test to verify the robustness of the SHHK Stock Connect's effect by assuming the SHHK Stock Connect launched two year later, which in November 2016. Columns (1)–(6) gradually incorporate firm-level control variables. The dependent variable is the monthly AH price premium. \texttt{FakePolicy} is a binary indicator equal to one for the period after November 2016 and zero otherwise. \texttt{Turnover}, \texttt{SOE}, \texttt{Demand}, \texttt{Size}, and \texttt{Interest Rate} are included as control variables. Endogenous variables are instrumented using their own lags dated $t-2$ and $t-3$. The $p$-values for the AR(2) and Hansen tests are provided to verify the absence of second-order serial correlation and the validity of the instruments, respectively. Firm and year fixed effects are included. Robust standard errors are reported in parentheses.
\item $^{***}$, $^{**}$, and $^{*}$ denote statistical significance at the 1\%, 5\%, and 10\% levels, respectively.
\end{tablenotes}
\end{threeparttable}
\end{sidewaystable}

\def\sym#1{\ifmmode^{#1}\else\(^{#1}\)\fi}
\begin{sidewaystable}[htbp]

\centering
\caption{Placebo test for Efficiency-Dependent Effects of the SHHK Stock Connect}
\label{tab:F2}
\setlength{\tabcolsep}{100pt}
\begin{threeparttable}
\begin{tabular}{l c}
\toprule
 & \multicolumn{1}{c}{AH Premium} \\
\cmidrule(lr){2-2}
 & (1) \\
\midrule
\texttt{Lagged Premium}      
    & \textbf{0.680\sym{***}} \\
    & (0.103) \\
\addlinespace

\texttt{FakePolicy} 
    & 0.150 \\
    & (0.127) \\
\addlinespace

\texttt{Effboth}     
    & -1.852 \\
    & (4.273) \\
\addlinespace

\texttt{FakePolicy} $\times$ \texttt{Effboth} 
    &  11.928 \\
    & (10.301) \\
\addlinespace

\texttt{Turnover}    
    & \textbf{0.019\sym{***}} \\
    & (0.006) \\
\addlinespace

\texttt{SOE}         
    & 0.026 \\
    & (0.042) \\
\addlinespace

\texttt{Demand}      
    & -0.007 \\
    & (0.009) \\
\addlinespace

\texttt{Size}       
    & \textbf{-0.080\sym{**}} \\
    & (0.032) \\
\addlinespace

\texttt{Interest rate}
    & -0.001 \\
    & (0.002) \\
\midrule

Observations&    6363         \\
AR(2) $p$-value&       0.466         \\
Hansen Test $p$-value&       0.849         \\
\bottomrule
\end{tabular}
\begin{tablenotes}
\footnotesize
\item \textit{Notes}: This table presents the results of a placebo test to verify the robustness of the efficiency-dependent effects of the SHHK Stock Connect. The dependent variable is the monthly AH price premium. To ensure the original findings are not driven by spurious factors, the implementation of the SHHK Stock Connect is assumed to have occurred in November 2016, two years after its actual launch. \texttt{FakePolicy} is a binary indicator equal to one for observations after November 2016 and zero otherwise. \texttt{Effboth} denotes the mutual market efficiency. The model is estimated using a two-step system GMM approach. Endogenous variables are instrumented using their own lags dated $t-2$ and $t-3$. The $p$-values for the AR(2) and Hansen tests are provided to verify the absence of second-order serial correlation and the validity of the instruments, respectively. Firm and year fixed effects are included. Robust standard errors are reported in parentheses.
\item $^{***}$, $^{**}$, and $^{*}$ denote statistical significance at the 1\%, 5\%, and 10\% levels, respectively.

\end{tablenotes}
\end{threeparttable}
\end{sidewaystable}

\newpage
\section{Robustness Check: Extra Control Variables}\label{app:G}

To ensure that our primary findings are not driven by the listing sequence of A-H dual-listed firms, we consider the listing sequence indicator, \texttt{AtoH}. It equals to one if a firm listed its A-shares prior to its H-shares, and zero otherwise. In our sample, 14 firms are listed first in Shanghai and subsequently in Hong Kong, 2 firms are listed simultaneously in both markets\footnote{The two simultaneously listed firms are China CITIC Bank International (A-share code: 601998.SH; H-share code: 0998.HK) and ICBC (A-share code: 601398.SH; H-share code: 1398.HK). These two firms are excluded in this part of the analysis.}, and the remaining 51 firms are initially listed in Hong Kong before cross-listing in Shanghai.

As reported in Column (2) of Table \ref{tab:G1}, the core results remain highly consistent with our baseline model. Specifically, the coefficient on the interaction term \texttt{SHHKPolicy $\times$ Effboth} remains negative and statistically significant at the 1\% level. This stability confirms that the policy's heterogeneous impact, conditioned on pre-existing market efficiency, is a robust phenomenon that persists regardless of the firm's listing path.

Interestingly, the coefficient of \texttt{AtoH} is negative and statistically significant at the 5\% level. This suggests that firms listing A-shares first generally exhibit a lower A-H premium compared to those that list H-shares first, holding other variables remaining constant. This finding aligns with the anchoring effect hypothesis, that is, for A-first firms, the pre-existing A-share price provides a transparent reference point for H-share institutional investors, thereby curbing speculative pricing and narrowing the initial valuation gap \citep{changCrosslistingPricingEfficiency}.

An alternative explanation is the regulation difference between A-share market and H-share market. Based on \citet{zhangMarketReactionCrossborder2022}, for A-to-H firms, prior listing in the Mainland China means that they have already passed a stricter domestic approval process than that required for listing in the Hong Kong stock market and have built a public disclosure record before entering Hong Kong. Once cross-listed in Hong Kong, these firms are priced in a market with a stronger information environment and lower information asymmetry, which supports more efficient price discovery. This channel helps explain why A-to-H firms tend to exhibit a lower AH premium.

We further consider the 2015 Chinese stock market turbulence to ensure that our main results are not driven by market-wide shocks. Specifically, we construct two indicator variables to capture different phases of the turbulence. The first, \texttt{turbulence2015wide}, equals one for the broader turbulence period from June 2015 to February 2016, and zero otherwise. The second, \texttt{turbulence2015narrow}, equals one for the core crash window from June 2015 to July 2015, and zero otherwise.

Table~\ref{tab:G2} reports the estimation results. In Column (1), the coefficient on \texttt{turbulence2015wide} is positive but statistically insignificant, suggesting that the whole turbulence period is not significantly associated with changes in the A-H share price premium. In Column (2), the coefficient on \texttt{turbulence2015narrow} is positive and statistically significant at the 5\% level, indicating that the A-H share price premium increased by about 9.5\% on average during the core crash window, holding other factors constant. Importantly, across both specifications, our key coefficients, \texttt{FakePolicy} $\times$ \texttt{Effboth}, remain statistically significant. 

In short, after controlling other potential factors that may affect our baseline results, the previous conclusion is still valid, indicating our previous baseline result is robust.

\def\sym#1{\ifmmode^{#1}\else\(^{#1}\)\fi}
\begin{sidewaystable}[htbp]
\centering
\caption{Controlling for A-H dual listing sequence bias}
\label{tab:G1}
\setlength{\tabcolsep}{60pt}
\begin{threeparttable}
\begin{tabular}{l*{2}{c}}
\toprule
            &\multicolumn{1}{c}{AH Premium}&\multicolumn{1}{c}{AH Premium}\\
            \cmidrule(lr){2-3}
            &\multicolumn{1}{c}{(1)}&\multicolumn{1}{c}{(2)}\\
\midrule
\texttt{Lagged Premium}      &       \textbf{0.696\sym{***}}&       \textbf{0.628\sym{***}}\\
            &     (0.139)         &     (0.081)         \\
\addlinespace
\texttt{AtoH}        &      -0.088         &      -0.093\sym{**} \\
            &     (0.060)         &     (0.047)         \\
\addlinespace
\texttt{SHHKPolicy}  &       \textbf{0.183\sym{***}}&      \textbf{-0.495\sym{**}} \\
            &     (0.022)         &     (0.197)         \\
\addlinespace
\texttt{Effboth}     &                     &      \textbf{39.952\sym{***}}\\
            &                     &    (14.989)         \\
\addlinespace
\texttt{SHHKPolicy $\times$ Effboth}    &                     &     \textbf{-46.036\sym{***}}\\
            &                     &    (14.502)         \\
\addlinespace
\texttt{Turnover}    &       \textbf{0.018\sym{**}} &       \textbf{0.024\sym{***}}\\
            &     (0.009)         &     (0.007)         \\
\addlinespace
\texttt{SOE}         &       0.008         &      -0.002         \\
            &     (0.047)         &     (0.040)         \\
\addlinespace
\texttt{Demand}      &       0.004         &       0.001         \\
            &     (0.013)         &     (0.012)         \\
\addlinespace
\texttt{Size}        &      -0.073         &      \textbf{-0.108\sym{***}}\\
            &     (0.044)         &     (0.031)         \\
\addlinespace
\texttt{Interest rate}&       0.002         &       \textbf{0.007\sym{***}}\\
            &     (0.002)         &     (0.002)         \\
\midrule
Observations&    6163         &    6163         \\
AR(2) $p$-value&       0.809         &       0.243         \\
Hansen Test $p$-value&       0.474         &       0.789         \\
\bottomrule
\end{tabular}
\begin{tablenotes}
\footnotesize
\item \textit{Notes}: This table reports the two-step system GMM estimation results investigating the impact of listing sequence and market efficiency on the AH price premium. The dependent variable is the monthly AH price premium. \texttt{AtoH} is a dummy variable equal to one for firms that listed on the Mainland (A-shares) prior to cross-listing in Hong Kong (H-shares), and zero otherwise. \texttt{SHHKPolicy} is a binary indicator equal to one for the period following the implementation of the Shanghai-Hong Kong Stock Connect (November 2014) and zero otherwise. \texttt{Effboth} denotes the mutual market efficiency. \texttt{Turnover}, \texttt{SOE}, \texttt{Demand}, \texttt{Size}, and \texttt{Interest rate} are included as control variables. Endogenous variables are instrumented using their own lags dated $t-2$ and $t-3$. The $p$-values for the AR(2) and Hansen tests are provided to verify the absence of second-order serial correlation and the validity of the instruments, respectively. Firm and year fixed effects are included. Robust standard errors are reported in parentheses.
\item $^{***}$, $^{**}$, and $^{*}$ denote statistical significance at the 1\%, 5\%, and 10\% levels, respectively.
\end{tablenotes}
\end{threeparttable}
\end{sidewaystable}

\def\sym#1{\ifmmode^{#1}\else\(^{#1}\)\fi}
\begin{sidewaystable}[htbp]

\centering
\caption{Controlling for the Chinese stock market turbulence shock}
\label{tab:G2}
\setlength{\tabcolsep}{80pt}
\begin{threeparttable}
\begin{tabular}{l*{2}{c}}
\toprule
            &\multicolumn{1}{c}{AH Premium}&\multicolumn{1}{c}{AH Premium}\\
            \cmidrule(lr){2-3}
            &\multicolumn{1}{c}{(1)}&\multicolumn{1}{c}{(2)}\\
\midrule
\texttt{Lagged Premium}      &   \textbf{0.578\sym{***}}&   \textbf{0.656\sym{***}}\\
            &     (0.110)          &     (0.073)          \\
\addlinespace
\texttt{SHHKPolicy}  &  \textbf{-0.522\sym{**}} &  \textbf{-0.476\sym{**}} \\
            &     (0.205)          &     (0.194)          \\
\addlinespace
\texttt{Effboth}     &  \textbf{46.127\sym{***}}&  \textbf{39.199\sym{***}}\\
            &    (16.812)          &    (14.634)          \\
\addlinespace
\texttt{SHHKPolicy $\times$ Effboth}    & \textbf{-50.532\sym{***}}& \textbf{-45.247\sym{***}}\\
            &    (15.574)          &    (14.233)          \\
\addlinespace
\texttt{Turnover}    &   \textbf{0.026\sym{***}}&   \textbf{0.021\sym{***}}\\
            &     (0.008)          &     (0.006)          \\
\addlinespace
\texttt{SOE}         &       0.024          &       0.021          \\
            &     (0.052)          &     (0.042)          \\
\addlinespace
\texttt{Demand}      &      -0.011          &      -0.008          \\
            &     (0.013)          &     (0.010)          \\
\addlinespace
\texttt{Size}        &  \textbf{-0.132\sym{***}}&  \textbf{-0.103\sym{***}}\\
            &     (0.044)          &     (0.029)          \\
\addlinespace
\texttt{Interest rate}&   \textbf{0.013\sym{***}}&   \textbf{0.009\sym{***}}\\
            &     (0.003)          &     (0.002)          \\
\addlinespace
\texttt{Turbulence2015wide}&       0.104          &                      \\
            &     (0.063)          &                      \\
\addlinespace
\texttt{Turbulence2015narrow}&                     &   \textbf{0.095\sym{**}} \\
            &                      &     (0.038)          \\
\midrule
Observations&    6363         &    6363         \\
AR(2) $p$-value&       0.136         &       0.322         \\
Hansen Test $p$-value&       0.887         &       0.845         \\
\bottomrule
\end{tabular}
\begin{tablenotes}
\footnotesize
\item \textit{Notes}: This table reports the two-step system GMM estimation results investigating the impact of Chinese stock turbulence happened in 2015 and market efficiency on the AH price premium. The dependent variable is the monthly AH price premium. \texttt{Turbulence2015wide} is an indicator equal to one for the period from June 2015 to February 2016, and zero otherwise. \texttt{Turbulence2015narrow} is an indicator to one for the period from June 2015 to July 2015, and zero otherwise. \texttt{SHHKPolicy} is a binary indicator equal to one for the period following the implementation of the Shanghai-Hong Kong Stock Connect (November 2014) and zero otherwise. \texttt{Effboth} denotes the mutual market efficiency. \texttt{Turnover}, \texttt{SOE}, \texttt{Demand}, \texttt{Size}, and \texttt{Interest rate} are included as control variables. Endogenous variables are instrumented using their own lags dated $t-2$ and $t-3$. The $p$-values for the AR(2) and Hansen tests are provided to verify the absence of second-order serial correlation and the validity of the instruments, respectively. Firm and year fixed effects are included. Robust standard errors are reported in parentheses.
\item $^{***}$, $^{**}$, and $^{*}$ denote statistical significance at the 1\%, 5\%, and 10\% levels, respectively.
\end{tablenotes}
\end{threeparttable}
\end{sidewaystable}

\end{document}